\documentclass[12pt]{article} 
\usepackage{axodraw}
\begin{document} 
\begin{titlepage}

\vspace*{2mm}

\begin{center} 
 
{\Large \bf 
XML-Based Formulation\\ 
of Field Theoretical Models\renewcommand{\thefootnote}{$\dagger$}
{\footnote{Talk presented by A.Kryukov at the International Workshop 
``Automatic Calculation for Future Colliders'' (CPP2001), 
November 28-30, 2001, Tokyo, Japan.}}\\}
\vspace{5mm}
{\large \bf \it A Proposal for a Future Standard
and Data Base for Model Storage, Exchange
and Cross-checking of Results\renewcommand{\thefootnote}{$\ddagger$}
{\footnote{Work is partially supported by CERN-INTAS 99-377, 
INTAS 00-0679 and RFBR 01-02-16209 grants.}}
} 
 
\vskip 1 cm 
 
{\bf A.Demichev, A.Kryukov and A.Rodionov}

\vskip 0.5cm

{\it Skobeltsyn Institute of Nuclear Physics,\\
     Moscow State University, 119992 Moscow, Russia}

\end{center} 
 
\begin{abstract}
\normalsize 
We propose an XML-based standard for formulation of field theoretical 
models. The goal of creation of such a standard is to provide a way for 
an unambiguous exchange and cross-checking of results of computer 
calculations in high energy physics. At the moment, the suggested 
standard implies that models under consideration are of the SM or 
MSSM type (i.e., they are just SM or MSSM, their submodels, 
smooth modifications or straightforward generalizations).
\end{abstract}

\end{titlepage}

\section{Introduction}
The progress of high energy experimental physics and the general 
interest in analysis of the Standard Model (SM) as well as its various 
modifications require the accurate theoretical computations of process 
characteristics to compare experimental results with theoretical 
predictions. The basic tool for this is the Feynman diagrammatic 
technique of computing matrix elements and, consequently, 
all physics quantities in high energy physics (HEP). However, as 
the number of final state particles grows (due to grow of the 
 beam energy of particle accelerators) the number of relevant 
Feynman diagrams becomes huge. The same is true for various generalizations 
of the Standard Model (even for relatively small number of final 
state particles), in particular, in the case of the Minimal Supersymmetric 
Standard Model (MSSM) because of large number of intermediate propagators. 
Therefore computation of all but the most simple processes is a quite 
lengthy task, prone to errors and mistakes. This means that Feynman 
amplitude calculation becomes impossible for practically interesting 
processes when one calculates cross sections by hands. 

Fortunately, the real power of the Feynman approach lies, through 
the use of definite rules, on {\it straightforward} transformation of each 
diagram in an algebraic expression representing its quantitative 
contribution to the process. Thus the perturbative calculation 
in quantum field theory can be realized as an {\it automatic computation 
system}.  There appeared several such systems, for instance, 
{\it CompHEP} \cite{CompHEP}, {\it Grace} \cite{Grace}, 
{\it MadGraph} \cite{MadGraph}, {\it VecBos} \cite{VecBos},  
{\it WbbGen} \cite{WbbGen}, and both theorists and experimentalists 
can benefit from these powerful packages for speeding up time consuming 
calculations. Another set of packages, such as {\it Herwig} \cite{Herwig},
{\it Isajet} \cite{Isajet}, {\it Pythia} \cite{Pythia} were developed for 
preparing event generations.

A typical {\it fully} automatic computing system for matrix element 
calculation has the following main modules: 
\begin{itemize}
\item
model definition;
\item
process definition;
\item
graph generation, drawing and selection;
\item
matrix element elaboration;
\item
analysis of kinematics and phase space integration;
\item
calculation of cross section;  
\item
event generation. 
\end{itemize}

An important condition for a successful use of automatic computation 
system in general, 
for cross-checking and verification of results obtained with the help 
of different packages as well as for a correct extraction of physical 
information is an unambiguous and transparent representation of 
the input data. The main goal of the present work is to develop and 
implement a more of less universal {\bf Standard for Model Formulations} 
(SMF) in order to provide:
\begin{itemize}
\item
an easy and unambiguous exchange through Internet by different 
QFT models between groups carrying out automatic computer calculations 
in high energy physics;
\item
a possibility for creation of interfaces automatically transforming 
a model presented in the standard into inputs for different automatic 
computing systems.
\end{itemize}

Our proposal is currently dealing only with the first stage of automatic 
calculations, namely, with a {\it model definition}. In the future, 
the standard should be extended to the second stage, i.e., 
{\it process definition}, as well.

Notice that at first sight it seems enough to fix the Lagrangian of 
a quantum field theoretical (QFT) model to define it completely - 
at least, at the perturbation theory 
level (in general, one should, in addition, fix boundary conditions). 
However, in the case of real computer calculations, the numerical results 
may, in general, depend on choice of a set of constants (e.g., coupling 
constants, masses, etc.) which are considered as a basic one (the obvious 
necessary condition for such a set of constants is that all other 
constants entering the model can be expressed through the basic ones), 
see, e.g. \cite{Gallix} and refs. therein. 
Another potential source of possible differences in results obtained 
for models with identical Lagrangians is different ways of representations 
of mathematical transcendental numbers like $\sqrt{2}$, $\pi$, etc. 
Therefore, the standard should include, in addition to Lagrangians, 
information about these peculiarities.

One more important requirement to the practical standard is an easy 
perception and understanding of a model. For this aim it should include 
elements of classification of models (e.g., submodel of SM, beyond 
MSSM, etc.), general description of the model, lists of fields and 
corresponding particles (in general, they are not in one-to-one 
correspondence: some fields  can be auxiliary and do not produce 
any particles). Thus from purely theoretical point of view, the 
standard contains exceeding information which, however, proves to be 
important for practical purposes.

\section{The Language for Representation of the Standard for Model 
Formulations}

For a successful development of the standard for model definition 
an appropriate method of representation and underling language 
must be chosen. To achieve the goals, the chosen language has to 
\begin{enumerate}
\item
be suitable and flexible enough for an adequate storage and representation 
of all the details needed for (perturbative) analysis of QFT models;
\item 
be platform independent and allow an easy computer processing of 
the model input data (set of fields, 
particles, their quantum numbers, vertices, etc.), in particular, 
a creation of an interface between the formulation of QFT models in this 
standard and computer programs for matrix element calculations or event 
generations;   
\item
provide an easy exchange of the data through computer nets, in particular, 
through the Internet;
\item
provide storage of the information about models {\it separately} from its 
visual representation: this essentially simplifies computer data 
processing, creation of the interface and allow an easy modification 
of the visual representation by users depending on their specific 
aims and tastes.  
\end{enumerate}
  
In our opinion, the best candidate for such a language is 
the {\bf XML} (Extensible Markup Language) which satisfies 
all the above requirements. The attractive features of XML which make 
it suitable for the development of SMF are the following 
(for an introduction to XML, see, e.g., \cite{XML,XML-D}):
\begin{itemize}
\item
XML has been designed to allow every meaningful division of 
a document to be unambiguously identified as part of a coherent 
tree structure that either a human or a computer can use. Thus, XML 
provides an application-independent format in which data can be shared.
\item
To achieve such an application-independence, this is for disparate 
groups may agree to build and use for their applications a 
defined Document Type Definition (DTD). 
The latter defines the structure of the documents. 
Groups emplying the same DTD then know that they can use data from 
applications created by any other groups. Moreover, there is the 
potential for exchange of data between parties without significant prior 
agreement (e.g., through WWW): a DTD sent with the XML data can provide 
the recipient with all the information they need to interpret and use it. 
\item 
The nature of XML data is not dependent on specific features of the 
platform on which it is used. Hence, if one upgrades and extends 
a system or application XML data can be still interpreted without 
requiring additions to the system foundations (like emulators). Especially 
this is important if one makes data or information publicly or widely 
accessible to others.
\item 
An essential advantage of XML is an existence of public 
libraries of parsers for XML-files suitable for development of 
various application program interfaces (API).   
\item
XML is, in fact, a meta-language, a special language that allow one to 
completely describe a class of other languages, which in turn describe 
documents.  Each of the latter languages are designed for every specific 
purpose to reach it in the most effective way.\\ 
{\it Example}: XML-based language MathML\cite{MathML} aimed at representation 
and transmission of mathematical expressions through the Internet.    
\item
With the help of special tools (Cascading Style Sheets (CSS), 
Extencible Stylesheet Language (XSL), etc) 
the information stored in XML-language can be represented in a visual 
form using the {\it standard Internet browsers} (such as 
Internet Explorer, Mozilla, Amaya, coming version of Netscape) 
similarly to the usual HTML-pages.  
\end{itemize}

\section{The Scope of the Standardization}

One may aim at different levels of systematization of QFT models and 
the corresponding standardization of their formulation.

Thanks to its generality and conciseness, 
the Feynman diagram method has spread, besides high energy physics, over 
many other research fields dealing with many-body systems like atomic, 
nuclear and solid state physics. Therefore, at the most general level 
it would be desirable to describe an arbitrary QFT model. 
In this case, models have to be appropriately classified, in particular,
according to the following general characteristics:
\begin{itemize}
\item
dimensionality of the space-time in which the model is formulated
(low dimensional models are interesting for testing new theoretical 
ideas and in the solid-body physics while higher dimensional 
models are nowadays seriously considered as promising candidates for 
solution of some long-standing problems in high energy physics);
\item
existence or absence of gauge invariance;
\item
existence of supersymmetry (including separation into component (on-shell) 
and superspace (off-shell) SUSY-formulations), {\it etc}.
\end{itemize}

Because of their generality this level requires, as a {\it starting} 
point, use of the corresponding Lagrangians in their {\it explicit 
form}. Then, using the information stored with the help of XML and MathML, 
a special computer program should produce the corresponding Feynman 
rules, also presented in the XML-format. At the moment, a 
realization of this project is in its very beginning.

As a first step, it is reasonable to develop the standard only for 
SM- or MSSM-like gauge models including their submodels (QED, QCD, etc.) 
as well as their closest modifications and generalizations. Moreover, 
in the present work we restrict ourselves to a simple example of the 
modified electrodynamics, namely the QED with additional four-fermion 
interactions described with the help of an auxiliary non-dynamical 
neutral field (that is the low-energy approximation for the interactions 
mediated by the Z-bosons).

\section{General Structure of the Standard}

The Standard starts from an acronym for the chosen model following by 
an indication of a type of the model (e.g., subset of modified SM; beyond 
MSSM; modified MSSM, etc.). Further information about a QFT model in 
the proposed standard is separated into ten parts:
\begin{enumerate}
\item {\bf General properties.}
This part contains the following general characteristics of models:
\begin{itemize}
\item[--] gauge group;
\item[--] existence of supersymmetry and number of SUSY generators;
\item[--] information about interactions and the corresponding quantum 
numbers in the model under consideration;
\item[--] existence or absence of Higgs particles;
\item[--] general features of the matter sector (e.g., number of 
generations);
\item[--] type of chosen gauge conditions.
\end{itemize}
\item {\bf Fields entering the model.} 
The second part contains information about quantum fields in the model: 
their physical meaning, properties under the Lorentz transformations, 
reality or complexity and chosen symbolic notation for them. 
\item {\bf Particles entering the model.} 
This part presents the corresponding particles together with 
their characteristics (mass, spin, etc.).
\item {\bf Basic physical constants.} 
Here the basic set of {\it independent} physical constants entering 
the model is fixed. In general, this set can be chosen in different 
ways and, in principle, the different sets may lead to apparent 
discrepancies in results for the same theoretical model.  
\item {\bf Dependent (auxiliary) physical constants.}
This part contains a list of important physical constants which 
can be expressed through the basic ones.
\item {\bf Way of computer representation of mathematical constants.}
Here a way for computer representation of transcendental mathematical 
constants (like, e.g., $\pi$, $\sqrt{2}$, etc.) is defined 
(it can be symbolic or numerical representation; in the latter case, 
the chosen precision of the representation should be indicated).
\item {\bf Free Lagrangian density.}
This is the explicit expression (the corresponding part of the source file 
is written with the help of MathML) of the free part of the 
Lagranginan for the model under consideration.
\item {\bf Gauge conditions (explicit expressions).} This part fixes 
gauge conditions. The gauge fields corresponding to different subgroups 
of the total gauge group may satisfy to different gauge conditions. 
Therefore, the part contains the list of the subgroups and the 
corresponding conditions.
\item {\bf Propogators.} The free Lagrangian together with the gauge 
conditions define propagators for the fields. Thus, strictly speaking, 
the information (an explicit form of the propagators) containing in 
this part is exceeding. But it proves to be convenient for practical 
purposes because propagators directly enter the Feynman rules and 
expressions corresponding to Feynman diagrams. On the other hand, a 
derivation of propagators from the Lagrangian and gauge conditions 
requires more or less lengthy and accurate calculations. Thus their 
explicit presentation makes easier a comparison of input data.
\item {\bf Interaction Lagrangian and vertices.} This part contains 
terms in the interaction Lagrangian of the model and the corresponding 
factors in matrix elements (according to the Feynman rules).
\end{enumerate}

In the next section we describe a relatively simple example of 
a QFT model presented in the framework of the proposed Standard (XML-SMF).

\section{A Simple Example of XML-SMF}

As an example of XML-SMF we shall consider Quantum Electrodynamics (QED) 
with additional four-fermion interactions, the latter being described 
with the help of an auxiliary $Z$-boson field with non-dynamical 
propagator.

According to the steps described in the preceding section,
the XML-based formulation comprises of ten parts. The corresponding 
XML-codes (including MathML parts representing mathematical expressions) 
are given in the Appendix. A special interface 
program provides transformation of the information stored in 
the XML-file in the form suitable for direct 
use of this model as input data for the CompHEP program\cite{CompHEP} 
designed for matrix element calculations at the tree level.

As we mentioned above (section 2, point 4), visual representation 
of the model formulation in the Standard should be separated from the 
content. In general, this can be achieved with the help of 
CSS (Cascading Style Sheets) or XSL (Extensible Stylesheet Language), 
see, e.g. \cite{XML,XML-D}. However, at present (to our best knowledge)
both this ways of visualization of an information stored in XML-form 
are incompatible with MathML. Therefore, at the moment we use  
for the visualization of XML-SMF the usual tags of XHTML inserted 
directly into the XML-files (thus violating the principle of 
separation of an information and its visualization). The combined 
XML/XHTML files can be rendered by the Mozilla browzer 
(MathML-enabled version: Mozilla 0.9.8, see \cite{Mozilla}). 
As a result, the visual representation of QED with four-fermion 
interactions looks as follows:

\vspace{1cm}

\hrule

\vspace{1cm}

\begin{center}
{\Large\bf QED+4F}
\end{center}

\begin{itemize}
\item {\large{\bf{\em Type}}: modified subset of SM}
\end{itemize}

\begin{center}
{\large\bf I. GENERAL PROPERTIES}
\end{center}

{\bf {\em Full Name}: Quantum Electrodynamics with 4-fermion interactions.}

\vspace{8mm}

{\bf Gauge Group and SUSY} 

\renewcommand{\arraystretch}{1.5} 
\begin{tabular}{|l|l|}      \hline 
Gauge Group & SUSY \\ \hline\hline
$U_{EM}(1)$& No\\ \hline
\end{tabular}

\newpage

{\bf Fundamental Interactions}

\begin{tabular}{|l|c|c|c|c|} \hline
Interactions & Submodel & Charge & Existence in the 
                                       & Comments\\ 
& & & current model & \\ \hline\hline
Strong & QCD & color & no & \\ \hline
Electromagnetic & QED & el. charge & yes & \\ \hline
 &  & flavor, &  & low-energy effective, \\
Weak& QFD & hypercharge & yes & 4-fermion, neutral \\
 & & & & currents \\ \hline
Gravity & GTR & energy & no & \\ \hline
Nonstandard & & & no & \\
interactions & & & & \\ \hline
\end{tabular}

\vspace{5mm}

{\bf Higgs Sector} 

\renewcommand{\arraystretch}{1.5} 
\begin{tabular}{|l|l|}      \hline 
Minimal SM/MSSM Higgs sector & Additional Higgs fields \\ \hline\hline
No& No\\ \hline
\end{tabular}

\vspace{5mm}

{\bf Matter Sector} 

\renewcommand{\arraystretch}{1.5} 
\begin{tabular}{|l|c|c|}      \hline 
Number of & Neutrino (existence and & Matter particles beyond\\ 
Generations & type & SM/MSSM \\ \hline\hline
1& No& No\\ \hline
\end{tabular}

\vspace{5mm}

{\bf Gauge Conditions} 

\renewcommand{\arraystretch}{1.5} 
\begin{tabular}{|l|c|c|}      \hline 
Gauge (Sub)Group & Condition & Existence of Ghosts \\ \hline\hline
$U(1)$ & Feynman& No\\ \hline
\end{tabular}

\vspace{5mm}

\begin{center}
{\large\bf II. FIELDS ENTERING THE MODEL}
\end{center}

\vspace{5mm}

{\bf Fields} 

\renewcommand{\arraystretch}{1.5} 
\begin{tabular}{|l|c|c|c|}      \hline 
Physical Type & Lorenz Type & Symbol & Comments\\ \hline\hline
electromagnetic, gauge $U(1)$, real & vector &$A_\mu$& \\ \hline
matter field & Dirac spinor& $\psi$&\\ \hline
auxiliary, nondynamical, real & vector& $Z_\mu$ &\\ \hline
\end{tabular}

\vspace{5mm}

\begin{center}
{\large\bf III. PARTICLES ENTERING THE MODEL}
\end{center}

\vspace{5mm}

{\bf Particles} 

\renewcommand{\arraystretch}{1.5} 
\begin{tabular}{|l|c|c|c|c|c|}      \hline 
Name & Symbol & Corresponding Field & Antiparticle& Spin/Helicity&
Mass\\ \hline\hline
photon & $\gamma$ & $A_\mu$ & $\gamma$ & 1& $m_\gamma=0$\\ \hline
electron & $e^-$& $\psi$& $e^+$& 1/2 & $m_e$ \\ \hline
\end{tabular}

\vspace{5mm}

\begin{center}
{\large\bf IV. BASIC PHYSICAL CONSTANTS}
\end{center}

\vspace{5mm}

{\bf Basic Constants} 

\renewcommand{\arraystretch}{1.5} 
\begin{tabular}{|l|c|}      \hline 
Physical meaning & Symbol \\ \hline\hline
electron charge & $g_e$\\ \hline
sine of the Salam-Weinberg angle & $\sin\theta$ \\ \hline
dimensionful parameter entering the propagator for the auxiliary 
$Z$-field & $M_Z$ \\ \hline
\end{tabular}

\vspace{5mm}

\begin{center}
{\large\bf V. DEPENDENT (AUXILIARY) PHYSICAL CONSTANTS}
\end{center}

\vspace{5mm}

{\bf Auxiliary Constants} 

\renewcommand{\arraystretch}{1.5} 
\begin{tabular}{|l|c|c|}      \hline 
Physical meaning & Symbol & Relation to others\\ \hline\hline
cosine of the Salam-Weinberg angle & $\cos\theta$ & 
                      $\sqrt{1-\sin\theta}$ \\ \hline
\end{tabular}

\vspace{5mm}

\begin{center}
{\large\bf VI. REPRESENTATION OF MATH CONSTANTS}
\end{center}

\vspace{5mm}

{\bf Math Constants} 

\renewcommand{\arraystretch}{1.5} 
\begin{tabular}{|l|c|}      \hline 
Symbol & Way of Representation \\ \hline\hline
$\pi$ & symbolic \\ \hline
$\sqrt{2}$ & symbolic \\ \hline

\end{tabular}

\vspace{5mm}

\begin{center}
{\large\bf VII. FREE LAGRANGIAN DENSITY}
\end{center}

\vspace{5mm}

$L_0=-\frac{1}{4}F_{\mu\nu}F^{\mu\nu} 
    + {\rm i}\bar\psi\gamma^\mu\partial_\mu\psi
    - m_e\bar\psi\psi - M_Z^2Z_\mu Z^\mu
$

\vspace{2mm}

$
F_{\mu\nu}=\partial_\nu A_\mu - \partial_\mu A_\nu
$

\vspace{5mm}

 \begin{center}
{\large\bf VIII. GAUGE CONDITIONS (EXPLICIT FORM) }
\end{center}

\vspace{5mm}

{\bf Gauge Conditions} 

\renewcommand{\arraystretch}{1.5} 
\begin{tabular}{|l|c|c|}      \hline 
Gauge Subgroup & Name of the Gauge Condition & Explicit Form\\ \hline\hline
$U_{EM}$ & Feynman & $\partial_\mu A^\mu=0$ \\ \hline
\end{tabular}

\vspace{5mm}

 \begin{center}
{\large\bf IX. PROPAGATORS }
\end{center}

\vspace{5mm}

{\bf Propagators} 

\renewcommand{\arraystretch}{1.5} 
\begin{tabular}{|l|c|c|}      \hline 
Fields & Math Expression & Diagram Element\\ \hline\hline
$\langle A_\mu A_\nu\rangle$ & 
$-\frac{{\rm i}g_{\mu\nu}}{k^2+{\rm i}\epsilon}$ & 
 \unitlength=1mm 
\linethickness{0.4pt} 
\begin{picture}(33.00,9.00)(0.00,4.00) 
\put(5.00,5.00){\circle*{1.00}} 
\put(20.00,5.00){\circle*{1.00}} 
\put(5.00,10.00){\makebox(0,0)[rc]{$\mu$}} 
\put(20.00,10.00){\makebox(0,0)[lc]{$\nu$}} 
\put(12.00,11.00){\makebox(0,0)[cc]{$k$}} 
\Photon(14.29,14.29)(57.14,14.29){2}{4} 
\end{picture}   \\[3mm] \hline

$\langle Z_\mu Z_\nu\rangle$ & 
$-\frac{{\rm i}g_{\mu\nu}}{M^2_Z}$ & 
\unitlength=1mm  
\linethickness{0.4pt} 
\begin{picture}(33.00,9.00)(0.00,4.00)\unitlength=1mm 
\put(5.00,5.00){\circle*{1.00}} 
\put(20.00,5.00){\circle*{1.00}} 
\put(5.00,10.00){\makebox(0,0)[rc]{$\mu$}} 
\put(20.00,10.00){\makebox(0,0)[lc]{$\nu$}} 
\put(12.00,11.00){\makebox(0,0)[cc]{$k$}} 
\Gluon(14.29,14.29)(57.14,14.29){4}{4} 
\end{picture} \\[3mm] \hline

$\langle \bar\psi\psi\rangle$ & 
$-{\rm i}\frac{\gamma_\mu k^\nu+m_e}{k^2-m^2_e+{\rm i}\epsilon}$ & 
\unitlength=1mm 
\linethickness{0.4pt} 
\begin{picture}(33.00,9.00)(0.00,4.00) 
\put(5.00,5.00){\circle*{1.00}} 
\put(20.00,5.00){\circle*{1.00}} 
\put(5.00,10.00){\makebox(0,0)[rc]{$i$}} 
\put(20.00,10.00){\makebox(0,0)[lc]{$j$}} 
\put(12.00,11.00){\makebox(0,0)[cc]{$k$}} 
\ArrowLine(14.29,14.29)(57.14,14.29) 
\end{picture} \\[3mm] \hline

\end{tabular}

 \begin{center}
{\large\bf X. INTERACTION LAGRANGIAN AND VERTICES }
\end{center}

\vspace{5mm}

{\bf Propagators} 

\renewcommand{\arraystretch}{1.5} 
\begin{tabular}{|c|c|c|}      \hline 
Term in the Interaction Lagrangian & 
Factor in Matrix Elements & Diagram Element\\ \hline\hline

$g_e\bar\psi\gamma^\mu A^\mu\psi$ &
${\rm i}g_e\gamma^\mu(2\pi)^4\delta(p_1-p_2-k)$ &

\unitlength=1mm 
\linethickness{0.4pt} 
\begin{picture}(33.00,20.00)(0.00,8.00) 
\put(5.00,10.00){\circle*{1.00}} 
\put(15.00,10.00){\circle*{1.00}} 
\put(22.00,17.00){\circle*{1.00}} 
\put(22.00,3.00){\circle*{1.00}} 
\put(5.00,15.00){\makebox(0,0)[cc]{$\mu$}} 
%\put(22.00,22.00){\makebox(0,0)[lc]{$i$}} 
%\put(22.00,8.00){\makebox(0,0)[lc]{$j$}} 
\Photon(14.29,28.57)(42.86,28.57){2}{3} 
\ArrowLine(42.86,28.57)(62.86,48.57) 
\ArrowLine(62.86,8.57)(42.86,28.57) 
\end{picture}\\[5mm] \hline
&& \\
$ \frac{g_e}{4\sin\theta\cos\theta}\bar\psi\Big[
\gamma^\mu(1-\gamma_5)$ &
$ \frac{g_e}{4\sin\theta\cos\theta}\Big[\gamma^\mu(1-\gamma_5)$ & \\
$-4\sin^2\theta\gamma^\mu\Big] Z_\mu\psi$ &
$-4\sin^2\theta\gamma^\mu\Big](2\pi)^4 $ & \\
& $ \times\delta(p_1-p_2-k)$ & \\
&&\unitlength=1mm 
\linethickness{0.4pt} 
\begin{picture}(33.00,00.00)(0.00,-5.00) 
\put(5.00,10.00){\circle*{1.00}} 
\put(15.00,10.00){\circle*{1.00}} 
\put(22.00,17.00){\circle*{1.00}} 
\put(22.00,3.00){\circle*{1.00}} 
\put(5.00,15.00){\makebox(0,0)[cc]{$\mu$}} 
%\put(22.00,22.00){\makebox(0,0)[lc]{$i$}} 
%\put(22.00,8.00){\makebox(0,0)[lc]{$j$}} 
\Gluon(14.29,28.57)(42.86,28.57){2}{3} 
\ArrowLine(42.86,28.57)(62.86,48.57) 
\ArrowLine(62.86,8.57)(42.86,28.57) 
\end{picture} \\ \hline

\end{tabular}

%%%%%%%%%%%%%%%%%%%%%%%%%%%%%%%%%%%%%%%%%%%%%%%%%%%%%%%%%%%%%%
\section{Conclusion} 
%%%%%%%%%%%%%%%%%%%%%%%%%%%%%%%%%%%%%%%%%%%%%%%%%%%%%%%%%%%%%%

XML provides a suitable basis for development of convenient 
and effective standard for QFT model formulation with possibility of  
exchange through Internet and reliable comparison of the results obtained 
by different groups and computer programs. 

Of course, much work has to be done yet for realization of 
the whole program of development of the universal, transparent, 
user-friendly  and widely accepted standard (XML-SMF) for QFT model 
formulation. Our current proposal is only a first step in this 
direction.

\vspace{5mm}

{\bf Acknowledgments} 
A.K. is grateful to Prof. Y.Kurihara for kind hospitality during the 
CPP-2001 conference.

\section*{Appendix: XML-codes for Formulation of the QED with 
Four-Fermion Interaction}

{\small

\begin{verbatim}

<?xml version="1.0" ?>

<!--       XML-standard for QFT models. 
          SAMPLE: QED with 4-fermion interaction.
                Version of 14.01.2002                        -->

<!DOCTYPE html PUBLIC "-//W3C//DTD XHTML 1.0 Strict//EN" "mathml.dtd">

<MODEL name="QED+4F" type="SM" variation="subset" modification="modified">


<!-- ****************************************************** -->
<!-- **** PART I. GENERAL PROPERTIES **************************** -->
<!-- ****************************************************** -->

<GEN_DESCRIPTION> Quantum Electrodynamics with 4-fermion
                                         interaction.

  <GaugeGroupAndSUSY>
      <GaugeGroup> <math xmlns="http://www.w3.org/1998/Math/MathML">
                   <msub><mi>U</mi><mi>EM</mi></msub>
                   <mo>(</mo>
                   <mn>1</mn>
                   <mo>)</mo>
                            </math> 
      </GaugeGroup>
      <SUSY> No </SUSY>
  </GaugeGroupAndSUSY>

  <Interactions>
     <StrongQCDcolor existence="no"/>
     <ElectromagneticQEDel_charge existence="yes"/>
     <WeakQFDflavor_hypercharge existence="yes"> 
           low-energy effective, 4-fermion, neutral currents 
      </WeakQFDflavor_hypercharge>
      <GravityGTRenergy existence="no"/>
      <NonstandardInteractions existence="no"/>
   </Interactions>

   <HiggsSector>
      <MinimalSM-MSSM existence="no"/>
      <AdditionalHiggses existence="no"/>
   </HiggsSector>

   <MatterSector>
       <NumberOfGenerations Num="1"/>
       <Neutrino existenceANDtype="no"/>
       <ExtraMatter existence="no"/>
   </MatterSector>

    <GaugeConditions>
       <GaugeSubgroup type="U(1)" ghosts="no"> Feynman
            </GaugeSubgroup>
    </GaugeConditions>

</GEN_DESCRIPTION>

<!-- ****************************************************** -->
<!-- **** PART II. FIELDS ENTERING THE MODEL ***************** -->
<!-- ****************************************************** -->

<FIELDS>

  <FIELD LorentzType="vector" id = "A-photon"> electromagnetic, 
                                               gauge U(1), real
      <SYMBOL> <math xmlns="&mathml;">
                <mi><mrow>
                   <msub>
                  <mi>A</mi>
                 <mi>&mu;</mi>
                   </msub>
                   </mrow></mi>
                   </math> </SYMBOL>

   </FIELD>

    <FIELD LorentzType="Dirac spinor" id = "psi"> matter field, complex
      <SYMBOL> <math xmlns="&mathml;">
               &psi;
                   </math> </SYMBOL>

   </FIELD>  

  <FIELD LorentzType="vector" id = "Z-boson"> auxiliary, nondynamical, real
      <SYMBOL> <math xmlns="&mathml;">
                <mi><mrow>
                   <msub>
                  <mi>Z</mi>
                 <mi>&mu;</mi>
                   </msub>
                   </mrow></mi>
                   </math> </SYMBOL>

   </FIELD>

</FIELDS>

<!-- ****************************************************** -->
<!-- **** PART III. PARTICLES ENTERING THE MODEL ***************** -->
<!-- ****************************************************** -->

 <PARTICLES>

   <PARTICLE name="photon" pdgID="22">
      <SYMBOL> 
        <math xmlns="&mathml;"> <mi><mrow>&gamma;</mrow></mi>
                   </math> </SYMBOL>

      <CorrFIELD><math xmlns="&mathml;">
                <mi><mrow>
                   <msub>
                  <mi>A</mi>
                 <mi>&mu;</mi>
                   </msub>
                   </mrow></mi>
                   </math> </CorrFIELD> 

       <AntiPARTICLE name="photon" pdgID="22"> 
               <math xmlns="&mathml;"> <mi><mrow>&gamma;</mrow></mi>
                                 </math> </AntiPARTICLE>

       <SPIN_or_HELICITY> 1 </SPIN_or_HELICITY>

       <MASS> <math xmlns="&mathml;">
                <mi><mrow>
                   <msub>
                  <mi>m</mi>
                 <mi>&gamma;</mi>
                   </msub>=0
                   </mrow></mi>
                   </math> 
      </MASS>

     </PARTICLE>

     <PARTICLE name="electron" pdgID="11">     
      <SYMBOL> 
        <math xmlns="&mathml;"> <mi><mrow>
                   <msup>
                  <mi>e</mi>
                 <mo>-</mo>
                   </msup></mrow></mi>
                   </math> </SYMBOL>

      <CorrFIELD><math xmlns="&mathml;">
                <mi><mrow>
                 <mi>&psi;</mi>
                   </mrow></mi>
                   </math> </CorrFIELD> 

       <AntiPARTICLE name="positron" pdgID="-11"> 
                   <math xmlns="&mathml;"> <mi><mrow>
                   <msup>
                  <mi>e</mi>
                 <mo>+</mo>
                   </msup></mrow></mi>
                   </math> </AntiPARTICLE>

       <SPIN_or_HELICITY> 1/2 </SPIN_or_HELICITY>

       <MASS> <math xmlns="&mathml;">
                <mi><mrow>
                   <msub>
                  <mi>m</mi>
                 <mi>e</mi>
                   </msub>
                   </mrow></mi>
                   </math> 
        </MASS>

     </PARTICLE>

</PARTICLES>

<!-- ****************************************************** -->
<!-- **** PART IV. BASIC PHYSICAL CONSTANTS ***************** -->
<!-- ****************************************************** -->

<BasicCONSTANTS>

  <BCONSTANT> 
     <PhysicalMeaning> electron charge </PhysicalMeaning>
     <SYMBOL> <math xmlns="&mathml;">
                <mi><mrow>
                   <msub>
                  <mi>g</mi>
                 <mi>e</mi>
                   </msub>
                   </mrow></mi>
                   </math>      </SYMBOL>
   </BCONSTANT>

   <BCONSTANT> 
     <PhysicalMeaning> sine of the Salam-Weinberg angle </PhysicalMeaning>
     <SYMBOL> <math xmlns="&mathml;">
                <mrow><mi>sin</mi><mi>&theta;</mi>
                   </mrow>
                   </math>      </SYMBOL>
    </BCONSTANT>

   <BCONSTANT> 
     <PhysicalMeaning> dimensionful parameter entering the propagator
                  for the auxiliary Z-field     </PhysicalMeaning>
     <SYMBOL> <math xmlns="&mathml;">
                <mi><mrow>
                   <msub>
                  <mi>M</mi>
                 <mi>Z</mi>
                   </msub>
                   </mrow></mi>
                   </math>       </SYMBOL>
    </BCONSTANT>

</BasicCONSTANTS>

<!-- ****************************************************** -->
<!-- **** PART V. DEPENDENT (AUXILIARY) PHYSICAL CONSTANTS ******* -->
<!-- ****************************************************** -->

<AuxCONSTANTS>

   <ACONSTANT> 
     <PhysicalMeaning> cosine of the Salam-Weinberg angle </PhysicalMeaning>
     <SYMBOL> <math xmlns="&mathml;">
                <mi><mrow>
                   <mi>cos</mi>
                 <mi>&theta;</mi>
                   </mrow></mi>
                   </math>      </SYMBOL>
     <RELATION_TO_OTHERS> <math xmlns="&mathml;">
                <msqrt>
                   <mrow>
                   <mn>1</mn>
                      <mo>-</mo>
                      <msup>
                      <mi>sin</mi><mn>2</mn>
                        </msup>
                         <mi>&theta;</mi>
                   </mrow></msqrt>
                   </math>          </RELATION_TO_OTHERS>    
    </ACONSTANT>

</AuxCONSTANTS>

<!-- ****************************************************** -->
<!-- **** PART VI. REPRESENTATIN OF MATH CONSTANTS ***************** -->
<!-- ****************************************************** -->

<MathCONSTANTS>

  <MCONSTANT> 
         <SYMBOL> <math xmlns="&mathml;">
               <mrow> <msqrt><mn>2</mn></msqrt></mrow>
                   </math>      </SYMBOL>
          <CompRepresentation> symbolic  </CompRepresentation>
   </MCONSTANT>

  <MCONSTANT> 
         <SYMBOL> <math xmlns="&mathml;">
                <mrow><mi>&pi;</mi></mrow>
                   </math>      </SYMBOL>
          <CompRepresentation> symbolic  </CompRepresentation>
   </MCONSTANT>

</MathCONSTANTS>

<!-- ****************************************************** -->
<!-- **** PART VII. FREE LAGRANGIAN DENSITY ***************** -->
<!-- ****************************************************** -->


 <FreeLagrangian higgs_shift = " ">

  <!-- the parameter higgs_shift may have values "before Higgs field 
shift", "after Higgs field shift" or blank_space                   -->

  <math xmlns="&mathml;"> 
     <mrow> <msub><mi>L</mi><mn>0</mn></msub>
     <mo>=</mo>
      <mo>-</mo>
      <mi></mi>
      <mfrac><mn>1</mn><mn>4</mn></mfrac>
      <msub><mi>F</mi><mi>&mu;&nu;</mi></msub>
      <msup><mi>F</mi><mi>&mu;&nu;</mi></msup>
      <mo> + </mo>
     <mi>i</mi> <mover><mi>&psi;</mi><mo>&OverBar;</mo></mover>
      <msup><mi>&gamma;</mi><mi>&mu;</mi></msup>
      <msub><mo>&PartialD;</mo><mi>&mu;</mi></msub><mi>&psi;</mi>
      <mo>-</mo>
      <msub><mi>m</mi><mi>e</mi></msub>
      <mover><mi>&psi;</mi><mo>&OverBar;</mo></mover><mi>&psi;</mi>
      <mo>-</mo>
      <msubsup><mi>M</mi><mi>Z</mi><mn>2</mn></msubsup>
      <msup><mi>Z</mi><mi>&mu;</mi></msup> 
      <msub><mi>Z</mi><mi>&mu;</mi></msub>
       </mrow>
     </math>

      <math xmlns="&mathml;">
        <mrow> 
        <msub><mi>F</mi><mi>&mu;&nu;</mi></msub>
        <mo>=</mo>
        <msub><mo>&PartialD;</mo><mi>&nu;</mi></msub>
        <msub><mi>A</mi><mi>&mu;</mi></msub>       
        <mo>-</mo>
         <msub><mo>&PartialD;</mo><mi>&mu;</mi></msub>
        <msub><mi>A</mi><mi>&nu;</mi></msub> 
        </mrow>
      </math>

</FreeLagrangian>

<!-- ****************************************************** -->
<!-- **** PART VIII. GAUGE CONDITIONS (EXPLICIT EXPR)  *** -->
<!-- ****************************************************** -->

<GaugeConditionsExplicitExpr>
    <GaugeSubgroup> 
                  <math xmlns="&mathml;">
                   <msub><mi>U</mi><mi>EM</mi></msub>
                   <mo>(</mo>
                   <mn>1</mn>
                   <mo>)</mo>
                            </math> 

       <GaugeCondition name = "Feynman"> 
                  
          <math xmlns="&mathml;">
           <mrow> 
           <msub><mo>&PartialD;</mo><mi>&mu;</mi></msub>
           <msup><mi>A</mi><mi>&mu;</mi></msup>       
           <mo>=</mo><mn>0</mn>
           </mrow>
          </math>
             
       </GaugeCondition>

     </GaugeSubgroup>
</GaugeConditionsExplicitExpr>

<!-- ****************************************************** -->
<!-- **** PART IX. PROPAGATORS ***************** -->
<!-- ****************************************************** -->

<PROPOGATORS>

  <PROPOGATOR FieldRef1 ="A-photon" vectIndex1 = "mu" 
              FieldRef2 ="A-photon" vectIndex2 = "nu" img="phot_prp.gif">
     <fields> <math xmlns="&mathml;"> <msub><mrow>
                 <mi>&#60;</mi>
                <mrow>
                   <msub>
                  <mi>A</mi>
                 <mi>&mu;</mi>
                   </msub>
                   </mrow>
                <mrow>
                   <msub>
                  <mi>A</mi>
                 <mi>&nu;</mi>
                   </msub>
                   </mrow><mi>&#62;</mi></mrow><mn>0</mn></msub>
                </math>
      </fields>
      <expression>
      <math xmlns="&mathml;"> 
       <mrow><mo>-</mo>
         <mfrac><mrow><mi>i</mi><msub><mi>g</mi>
                <mi>&mu;&nu;</mi></msub></mrow>
                <mrow><msup><mi>k</mi><mn>2</mn></msup>
                <mo>+</mo><mi>i</mi><mi>&epsilon;</mi></mrow>
          </mfrac>
        </mrow>  
                     </math>
      </expression>
  </PROPOGATOR>
       
  <PROPOGATOR FieldRef1 ="psi" spinIndex1 = "i" 
              FieldRef2 ="psi" spinIndex2 = "j" img="el_prp.gif">
     <fields> <math xmlns="&mathml;"> <msub><mrow>
                 <mi>&#60;</mi>
                <mrow>
                   <mi>&psi;</mi>
                 <mi>&psi;</mi>
                      </mrow>
                <mi>&#62;</mi></mrow><mn>0</mn></msub>
                </math>
      </fields>
      <expression>
         <math xmlns="&mathml;"> 
            <mrow><mi>i</mi>
         <mfrac><mrow><msub><mi>&gamma;</mi><mi>&mu;</mi></msub> 
                 <msup> <mi>k</mi><mi>&mu;</mi></msup>
                  <mo>+</mo><msub><mi>m</mi><mi>e</mi></msub></mrow>
                <mrow><msup><mi>k</mi><mn>2</mn></msup>
                <mo>-</mo><msubsup><mi>m</mi><mi>e</mi><mn>2</mn></msubsup>
                <mo>+</mo><mi>i</mi><mi>&epsilon;</mi></mrow>
          </mfrac>
        </mrow>             
          </math>
      </expression>
  </PROPOGATOR>
 
  <PROPOGATOR FieldRef1 ="Z-boson" vectIndex1 = "mu" 
              FieldRef2 ="Z-boson" vectIndex2 = "nu" img="z_prp.gif">
     <fields> <math xmlns="&mathml;"> <msub><mrow>
                 <mi>&#60;</mi>
                <mrow>
                   <msub>
                  <mi>Z</mi>
                 <mi>&mu;</mi>
                   </msub>
                   </mrow>
                <mrow>
                   <msub>
                  <mi>Z</mi>
                 <mi>&nu;</mi>
                   </msub>
                   </mrow><mi>&#62;</mi></mrow><mn>0</mn></msub>
                </math>
      </fields>
      <expression>
      <math xmlns="&mathml;"> 
       <mrow><mo>-</mo>
         <mfrac><mrow><mi>i</mi><msub><mi>g</mi>
                <mi>&mu;&nu;</mi></msub></mrow>
                <mrow><msubsup><mi>M</mi><mi>Z</mi><mn>2</mn></msubsup>
                </mrow>
          </mfrac>
        </mrow>        
        </math>
      </expression>
  </PROPOGATOR>
 
</PROPOGATORS>

<!-- ****************************************************** -->
<!-- **** PART X. INTERACTION LAGRANGIAN AND VERTICES ***** -->
<!-- ****************************************************** -->

<VERTICES>

  <VERTEX vertexID="eep" img="eep.gif">

      <TERMinINTERACTION_LAGRANGIAN>
      <math xmlns="&mathml;"> 
          <mrow><msub><mi>g</mi><mi>e</mi></msub>
               <mover><mi>&psi;</mi><mo>&OverBar;</mo></mover>
          <msup><mi>&gamma;</mi><mi>&mu;</mi></msup>
          <msub><mi>A</mi><mi>&mu;</mi></msub><mi>&psi;</mi>
          </mrow>
        </math>
        </TERMinINTERACTION_LAGRANGIAN>

      <FACTORinMATRIX_ELEMENTS>
          <math xmlns="&mathml;"> 
           <mrow>       
            <mi>i</mi><msub><mi>g</mi><mi>e</mi></msub>
            <msup><mi>&gamma;</mi><mi>&mu;</mi></msup>
            <msup><mrow><mo>(</mo><mn>2</mn><mi>&pi;</mi><mo>)</mo></mrow> 
                  <mn>4</mn></msup>
            <mi>&delta;</mi><mo>(</mo><msub><mi>p</mi><mi>1</mi></msub>
                   <mo>-</mo><msub><mi>p</mi><mi>2</mi></msub>
                   <mo>-</mo><mi>k</mi><mo>)</mo>
           </mrow>
           </math>
          </FACTORinMATRIX_ELEMENTS>
  </VERTEX> 

  <VERTEX vertexID="eez" img="eez.gif">>

      <TERMinINTERACTION_LAGRANGIAN>
                 <math xmlns="&mathml;"> 
       <mrow>
       <mfrac><msub><mi>g</mi><mi>e</mi></msub>
       <mrow><mn>4</mn><mrow><mi>sin</mi><mi>&theta;</mi></mrow>
                <mrow><mi>cos</mi><mi>&theta;</mi></mrow></mrow></mfrac>
               <mover><mi>&psi;</mi><mo>&OverBar;</mo></mover>
          <mo>[</mo>
          <msup><mi>&gamma;</mi><mi>&mu;</mi></msup>
          <mo>(</mo><mn>1</mn><mo>-</mo> 
          <msub><mi>&gamma;</mi><mi>5</mi></msub><mo>)</mo>
          <mo>-</mo><mn>4</mn>
          <msup><mi>sin</mi><mi>2</mi></msup><mi>&theta;</mi> 
          <msup><mi>&gamma;</mi><mi>&mu;</mi></msup>
          <mo>]</mo>
          <msub><mi>Z</mi><mi>&mu;</mi></msub><mi>&psi;</mi>
          </mrow>
          </math>
        </TERMinINTERACTION_LAGRANGIAN>

      <FACTORinMATRIX_ELEMENTS>
                 <math xmlns="&mathml;"> 
            <mrow>       
            <mi>i</mi>
          <mfrac><msub><mi>g</mi><mi>e</mi></msub>
       <mrow><mn>4</mn><mrow><mi>sin</mi><mi>&theta;</mi></mrow>
                <mrow><mi>cos</mi><mi>&theta;</mi></mrow></mrow></mfrac>
           <mo>[</mo>
          <msup><mi>&gamma;</mi><mi>&mu;</mi></msup>
          <mo>(</mo><mn>1</mn><mo>-</mo> 
          <msub><mi>&gamma;</mi><mi>5</mi></msub><mo>)</mo>
          <mo>-</mo><mn>4</mn>
          <msup><mi>sin</mi><mi>2</mi></msup><mi>&theta;</mi> 
          <msup><mi>&gamma;</mi><mi>&mu;</mi></msup>
          <mo>]</mo>
          <msup><mrow><mo>(</mo><mn>2</mn><mi>&pi;</mi><mo>)</mo></mrow> 
           <mn>4</mn></msup>
            <mi>&delta;</mi><mo>(</mo><msub><mi>p</mi><mi>1</mi></msub>
                   <mo>-</mo><msub><mi>p</mi><mi>2</mi></msub>
                   <mo>-</mo><mi>k</mi><mo>)</mo>
           </mrow>
</math>
          </FACTORinMATRIX_ELEMENTS>
  </VERTEX> 
</VERTICES>
<!-- ******************************************************* -->
<!-- ******************************************************* -->
</MODEL>

\end{verbatim}
}

\begin{thebibliography}{99}
\bibitem{CompHEP}
A.~Pukhov {\it et al.},
%``CompHEP: A package for evaluation of Feynman diagrams and
%integration over multi-particle phase space. User's manual for 
%version 33,''
hep-ph/9908288.

\bibitem{Grace}
T.~Ishikawa, T.~Kaneko, K.~Kato, S.~Kawabata, Y.~Shimizu and H.~Tanaka
                  [MINAMI-TATEYA group Collaboration],
%``GRACE manual: Automatic generation of tree amplitudes in
% Standard Models: Version 1.0,''
KEK-92-19.

\bibitem{MadGraph}
T.~Stelzer and W.~F.~Long,
%``Automatic generation of tree level helicity amplitudes,''
Comput.\ Phys.\ Commun.\  {\bf 81}, 357 (1994)
[hep-ph/9401258].

\bibitem{VecBos}
F.~A.~Berends, H.~Kuijf, B.~Tausk and W.~T.~Giele,
%``On the production of a W and jets at hadron colliders,''
Nucl.\ Phys.\ B {\bf 357}, 32 (1991).

\bibitem{WbbGen}
F.~Caravaglios, M.~L.~Mangano, M.~Moretti and R.~Pittau,
%``A new approach to multi-jet calculations in hadron collisions,''
Nucl.\ Phys.\ B {\bf 539}, 215 (1999)
[hep-ph/9807570].

\bibitem{Herwig}
G.~Corcella {\it et al.},
%``HERWIG 6: An event generator for hadron emission reactions 
%with  interfering gluons (including supersymmetric processes),''
JHEP {\bf 0101}, 010 (2001)
[hep-ph/0011363].

\bibitem{Isajet}
H.~Baer, F.~E.~Paige, S.~D.~Protopopescu and X.~Tata,
%``ISAJET 7.48: A Monte Carlo event generator for p p, anti-p p, and  
% e+ e- reactions,''
hep-ph/0001086.

\bibitem{Pythia}
T.~Sjostrand, P.~Eden, C.~Friberg, L.~Lonnblad, G.~Miu, 
S.~Mrenna and E.~Norrbin,
%``High-energy-physics event generation with PYTHIA 6.1,''
Comput.\ Phys.\ Commun.\  {\bf 135}, 238 (2001)
[hep-ph/0010017].
\bibitem{Gallix}
D.~Perret-Gallix, ``Towards a Complete Feynman Diagrams 
Automatic Computation System'', hep-ph/9508235.
\bibitem{XML}
A.~Homer, {\it XML IE5 Programmer's Reference}, Wrox Press, 
Birmingham, 1999\\
\bibitem{XML-D}
H.M.~Deitel {\it et al.} {\it XML. How to Program}, Prentice Hall, 
Upper Saddle River, NJ, 2000
\bibitem{MathML} See the official Web-site on MathML {\tt 
http://www.w3.org/Math} 
\bibitem{Mozilla} See the Mozilla Web-site {\tt 
http://www.mozilla.org/projects/mathml/build.html}
\end{thebibliography}
\end{document}